\documentclass[%
    reprint, 
    superscriptaddress,
    nofootinbib,
    amsmath,amssymb,
    aps,
    prx,
    floatfix,
]{revtex4-2}

\usepackage{graphicx}
\usepackage{dcolumn}
\usepackage{bm}
\usepackage{xcolor}
\usepackage[normalem]{ulem}
\usepackage[english]{babel}
\usepackage[colorlinks=true,linkcolor=blue,citecolor=blue]{hyperref}
\usepackage{float}
\usepackage{siunitx}

\usepackage{physics}
\usepackage{textcomp}
\usepackage{siunitx}

\usepackage{verbatim} 
\usepackage{color}
\usepackage{subfigure} 
\usepackage{hyperref}
\usepackage{xr}
\usepackage{soul}
\usepackage{ulem}
\setstcolor{red}

\usepackage{enumitem}
\newlist{outline}{enumerate}{2}
\setlist[outline]{label=\Alph*.,noitemsep}
\setlist[outline,1]{label=\Alph*.}
\setlist[outline,2]{label=\arabic*.,leftmargin=*}
\setlist[outline,3]{label=\arabic*.}

\usepackage[
    textwidth=0.8in,
    textsize=footnotesize,
    color=red!25,
    colorinlistoftodos,
]{todonotes}

\begin{document}
\preprint{AIP/123-QED}

\title[Chemical Potential]{Engineering a multi-level bath for transmons with three-wave mixing and parametric drives}

\author{Xi Cao$^{*, \dagger}$}
\author{Maria Mucci$^*$}
\author{Gangqiang Liu}
\author{David Pekker}
\author{Michael Hatridge$^{\ddagger}$}
\affiliation{Department of Physics and Astronomy, University of Pittsburgh, Pittsburgh, PA, 15260, USA}

\date{May 4, 2025}

\begin{abstract}
A quantum system with a tunable bath temperature provides an additional degree of freedom for quantum simulators.
Such a system can be realized by parametrically modulating the coupling between the system and the bath.
Here, by coupling a transmon qubit to a lossy Superconducting Nonlinear Asymmetric Inductive eLement (SNAIL) mode, we experimentally create a tunable bath for the qubit mode.
The effective temperature of this bath can be precisely controlled, ranging from negative to positive values.
We show that the qubit can be thermalized to equilibrium with different population distributions under different parametric pumping conditions. 
We further extend our method to the third level of the transmon, demonstrating its potential utility beyond the two-level case. 
Our results provide a useful tool that can be readily integrated with quantum simulators that would benefit from a nontrivial photon population distribution. 
\end{abstract}

\maketitle

\def\thefootnote{*}\footnotetext{These authors contributed equally to this work}
\def\thefootnote{$\dagger$}\footnotetext{Contact author: xicao@illinois.edu; Now at: Department of Physics, University of Illinois Urbana-Champaign, Champaign, IL, USA}
\def\thefootnote{$\ddagger$}\footnotetext{Contact author: hatridge@pitt.edu}

\textit{Introduction} - A quantum simulator~\cite{FEYNMAN_Simulating_1982} can address complex problems in various domains of physics by mapping problems that are otherwise difficult to probe or track in their original settings to an engineered quantum system over which experimenters have good control. 
Such simulators have provided insights into a wide range of fields, such as quantum materials simulation~\cite{LEWENSTEIN_Ultracold_2007, GINGRAS_Quantum_2014}, quantum chemistry~\cite{REIHER_Elucidating_2017}, quantum transport~\cite{KRINNER_Twoterminal_2017, GIAZOTTO_Opportunities_2006}, many-body dynamics~\cite{ABANIN_Colloquium_2019, HARPER_Topology_2020}, and black holes~\cite{SWINGLE_Unscrambling_2018}.
Interactions between quantum systems and their uncontrolled environments are ubiquitous on different quantum simulation platforms, including superconducting microwave circuits~\cite{BLAIS_Circuit_2021}, trapped ions~\cite{MONROE_Programmable_2021}, photonic quantum machines~\cite{SLUSSARENKO_Photonic_2019}, and neutral atoms~\cite{HENRIET_Quantum_2020}.
These interactions are usually viewed as detrimental, as they lead to the decoherence of the quantum systems.
However, when engineered controllably, they tailor the properties of the environment seen by quantum systems, providing extra degrees of freedom for quantum simulations~\cite{HARRINGTON_Engineered_2022}.
Temperature, as one of the fundamental properties of the environment, affects a quantum system by influencing its thermodynamics, state occupation, decoherence, and transitions. 
Therefore, the ability to engineer an environment with a temperature that can be adjusted from negative to positive values is of great interest~\cite{shabani_artificial_2016}.
It enables the thermodynamical simulation of many-body systems~\cite{GEORGESCU_Quantum_2014}, allows the preparation of the Gibbs state~\cite{TRUSHECHKIN_Open_2022}, and provides an important knob for quantum annealing~\cite{DAS_Quantum_2005}.
So far, such systems have been realized in localized spin systems~\cite{PURCELL_Nuclear_1951, OJA_Nuclear_1997, MEDLEY_Spin_2011} and cold atoms in an optical lattice~\cite{BRAUN_Negative_2013}. 
In this letter, we create an engineered bath with a tunable temperature in the microwave domain.

Superconducting circuits and circuit quantum electrodynamics (cQED)~\cite{BLAIS_Circuit_2021} provide a promising platform for this task.
Bath engineering has become a powerful tool in cQED systems due to the strong coupling between superconducting circuits and their environment, as well as the controllable parametric interactions between different circuit elements~\cite{HARRINGTON_Engineered_2022}. 
In particular, it has been used to manipulate quantum systems, for example, to fast reset a qubit to the ground state with appropriate drives~\cite{GEERLINGS_Demonstrating_2013}, to stabilize a qubit to an arbitrary state~\cite{LU_Universal_2017}, and to stabilize a many-body system with a nontrivial state~\cite{MA_Dissipatively_2019}.

Here, we experimentally demonstrate a temperature-tunable bath for a transmon qubit.
We start by parametrically coupling the qubit to a lossy mode. 
For the first experiments, the transmon is treated as a two-level system whose dissipation rates are engineered by tuning the strengths of these two parametric processes.
We show that in the steady-state limit, the qubit reaches the Gibbs state: $\rho = e^{-\beta H}/\text{Tr}(e^{-\beta H})$, where $k_{B}$ is the Boltzmann constant, and $\beta = 1/ k_B T$ is the inverse temperature set by the dissipation rates.
We verify this behavior by stabilizing the qubit to various steady states with different population distributions, including a population-inverted (negative temperature) state, which cannot be achieved by coupling to a natural bath at positive temperatures.

\begin{figure*}[th]
\includegraphics{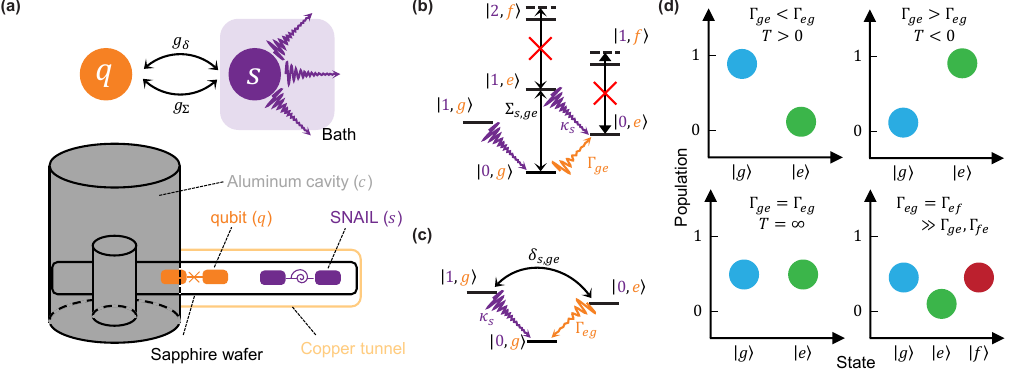}
\caption{\textbf{Experiment concept and schematic.} 
(a) Experiment setup.
The qubit and SNAIL are fabricated on the same sapphire chip, which is housed in a two-part enclosure comprising the aluminum cavity and a copper tunnel (see Fig.~\ref{fig:devicephoto} for a photograph of the device).
By applying external pump signals, different parametric interactions with rates $g_{\delta}$ and $g_{\Sigma}$ can be established between the qubit and SNAIL (see text for details).
(b), (c) Level diagram of the system. 
(b) Heating process with the $\Sigma$ drive. 
The system starts in its joint ground state and is brought to $\ket{1, e}$ with the $\Sigma$ pump. 
It will quickly falls into the $\ket{0, e}$ state due to the lossy nature of the SNAIL, which results in an effective ``heating'' process with rate $\Gamma_{ge}^p$.
(c) Cooling process with the $\delta$ drive. 
The qubit is prepared in $\ket{e}$ and the $\delta_{s, ge}$ drive converts the excitation to the SNAIL, which quickly decays. 
This process creates an effective parametric cooling rate $\Gamma_{eg}^p$ to empty the qubit.
(d) Tuning the artificial temperature with parametric processes. 
Positive temperature (top left) is achieved by setting $\Gamma_{eg} > \Gamma_{ge}$ while negative temperature (top right) is achieved with $\Gamma_{eg} < \Gamma_{ge}$. 
When the two rates are balanced, the system thermalizes to an infinite artificial temperature (bottom left). 
`Non-thermal' states where the transmon relaxes away from $\ket{e}$ without preference between $\ket{g}$ and $\ket{f}$ (bottom right).
}
\label{fig:experimentdiagram}
\end{figure*}

Furthermore, the anharmonic, multi-level nature of the transmon allows us to go beyond the simple two-level model.
We extend our parametric bath engineering technique to the third energy level of the transmon, creating a nontrivial steady state beyond the Gibbs state.
Accessing the higher levels of the transmon extends its utility as a tunable bath when coupling to a larger quantum simulator, and it also provides a potential tool to enter the regime of Gentile statistics~\cite{GENTILEJ._ItOsservazioni_1940, DAI_Gentile_2004, SRIVATSAN_Gentile_2006}, a generalization of Fermi-Dirac and Bose-Einstein statistics where the mode maximum occupation number (here, photons in a qubit) can be set between one (Fermions) and infinity (Bosons).

\textit{Model} - We begin by constructing a model that reflects our objective: create a temperature-tunable bath for a transmon qubit. 
Inspired by a proposal from Hafezi \textit{et al.}~\cite{HAFEZI_Chemical_2015}, this can be realized by parametrically coupling the qubit to another lossy mode. 
We consider a system that consists of a transmon~\cite{KOCH_Chargeinsensitive_2007} capacitively coupled to a SNAIL mode~\cite{FRATTINI_3wave_2017} as shown in the top panel of Fig.~\ref{fig:experimentdiagram}(a).
The SNAIL is used both as a three-wave mixing element and as the lossy mode. 
The qubit is also coupled to a 3D cavity for dispersive measurement~\cite{BLAIS_Cavity_2004}.
We note that while our experiment is implemented with cQED, the general model of the design is fundamentally platform-independent.

Central to our protocol are the parametric interactions between the qubit and the SNAIL.
In particular, we are interested in two types of interactions: two-mode squeezing and mode conversion, whose Hamiltonians are given by~\cite{ROY_Introduction_2016, KAMAL_Signalpump_2009, BERGEAL_Phasepreserving_2010}: 
\begin{align}
    H_{\text{tms}} &= g_{\scriptscriptstyle \Sigma} (s^{\dagger} q^{\dagger} + \text{h.c.}), \label{Eq: tms eq} \\
    H_{\text{bs}} &= g_{\scriptscriptstyle \delta} (s^{\dagger} q + \text{h.c.}), \label{Eq: bs eq} 
\end{align}
where $s$ and $q$ are the annihilation operators of the SNAIL and qubit mode, respectively; $g_{\scriptscriptstyle \Sigma}$ and $g_{\scriptscriptstyle \delta}$ are the interaction strength of the corresponding processes.
The subscripts are labeled $\Sigma$ and $\delta$ due to a pump at summation or difference frequency of the SNAIL ($\omega_\text{s}$) and qubit ($\omega_\text{q}$), i.e., $\omega_\text{p} = \omega_\text{s} \pm \omega_\text{q}$, which is required for each respective process (see Appendix~\ref{FullSystemHamiltonian} for details).

Equations~\ref{Eq: tms eq} and~\ref{Eq: bs eq} give a general description of the two processes between the linear modes.
In our case, however, the anharmonicity of the transmon qubit needs to be considered. 
Let us take the two-mode squeezing case as an example.
As shown in Fig.~\ref{fig:experimentdiagram}(b), the two-mode squeezing pump creates a coherent process between $\ket{0, g}$ and $\ket{1, e}$ (the joint ket notation follows the convention of $\ket{\text{SNAIL} (\text{number}) , \text{qubit} (\text{letter})}$). 
When both modes are linear, the pump will also drive similar processes, such as $\ket{1, e} \leftrightarrow \ket{2, f}$, $\ket{2, f} \leftrightarrow \ket{3, h}$, etc.
However, the transmon's large anharmonicity makes the pump off-resonant for these processes and thus prevents (at least to the leading order) them from appearing in the rotating frame.
It is then necessary to rewrite Eqs.~\ref{Eq: tms eq} and~\ref{Eq: bs eq} for individual qubit levels. 
For two adjacent qubit levels $\ket{i}$ and $\ket{i + 1}$, the respective interaction Hamiltonian terms are given by:
\begin{align}
    H_{\Sigma_{s, i, i+1}} &= g_{\scriptscriptstyle \Sigma_{s, i, i+1}}  s^{\dagger} \ket{i+1}\bra{i} + \text{h.c.}, \label{Eq: sigma pump Hamiltonian} \\ 
    H_{\delta_{s, i, i+1}} &= g_{\scriptscriptstyle \delta_{s, i, i+1}}  s^{\dagger} \ket{i}\bra{i+1} + \text{h.c.}, \label{Eq: delta pump Hamiltonain} 
\end{align}
where $g_{\scriptscriptstyle \Sigma_{s, i, i+1}}$ and $g_{\scriptscriptstyle \delta_{s, i, i+1}}$ are the effective coupling rates of the two individual processes (Appendix~\ref{parametricbathengineering}), with the corresponding pump frequencies: $\omega_{\scriptscriptstyle \Sigma_{s,i, i+1}}  = \omega_s + \omega_{i, i+1}$ and $\omega_{\scriptscriptstyle \delta_{s,i, i+1}} = \omega_s - \omega_{i, i+1}$.
We have chosen to replace the usual transmon raising/lowering operators in these equations with ket-bra notation to make clear that each parametric drive frequency links only a single pair of transmon levels.
Because of the frequency-selective nature of the parametric processes, multiple parametric interactions can be activated at the same time by simultaneously applying multiple pumps. 

Parametrically coupling the transmon to a lossy mode via these interactions provides an essential tool for creating the artificial temperature.
Let us start with the case where the transmon is treated as a two-level system.
As shown in Fig.~\ref{fig:experimentdiagram}(b), a pump at $\omega_p = \omega_{\scriptscriptstyle \Sigma_{s, ge}}$ coherently brings the system from $\ket{0, g}$ to $\ket{1, e}$.
When the SNAIL decay rate is much larger than the other relevant rates (i.e., $\kappa_s \gg g_{\scriptscriptstyle \Sigma_{s, ge}}, 1/\text{T}^\text{q}_1$), the system quickly falls into $\ket{0, e}$, and is effectively ``heated'' to a higher energy state.
Similarly, as shown in Fig.~\ref{fig:experimentdiagram}(c), a pump at $\omega_p = \omega_{\scriptscriptstyle \delta_{s, ge}}$ will effectively ``cool'' the transmon to its ground state. 

The tunability of this artificial temperature is realized by applying both pumps simultaneously.
By controlling the effective rates of the two processes, we can thermalize the transmon to an arbitrary equilibrium state, as shown in Fig.~\ref{fig:experimentdiagram}(d).
We show that (Appendix~\ref{chempot_derivation}) the density matrix of the equilibrium state can be described by a canonical ensemble:
\begin{equation}
\rho_{\text{q}} = e^{-\beta_{\text{q}} \hbar \omega_{\text{q}} \sigma^z_{\text{q}}/2}/\text{Tr}(e^{-\beta_{\text{q}} \hbar \omega_{\text{q}} \sigma^z_{\text{q}}/2})    
\label{Eq: CP rho}
\end{equation}
where the inverse temperature $\beta_{\text{q}}$ is given by:
\begin{equation}
\beta_{\text{q}} = \frac{1}{\omega_{\text{q}}} \text{ln} (\frac{ g_{\Sigma}^2  + g_{\delta}^2 e^{\beta_{\text{B}} \omega_{\text{s}} } }{g_{\Sigma}^2 e^{\beta_{\text{B}} \omega_{\text{s}} } + g_{\delta}^2 }).
\label{Eq: CP mu}
\end{equation}
This is a Gibbs state for a single qubit, whose temperature is a function of the controllable interaction strengths, $g_{\Sigma}$ and $g_{\delta}$.
In principle, this bath engineering technique can be extended to prepare the Gibbs state at an arbitrary temperature for a more generalized system (Appendix \ref{general Gibbs state}). 

Before moving on to the experiment results, we point out an important caveat.
Since each decay event is heralded by a photon in the SNAIL mode, we must take into account the effects of photon distinguishability when several processes create excitations at the same time. 
For example, some engineered bosonic decay processes~\cite{GERTLER_Protecting_2021} require photon indistinguishability, as the decay operator should protect a manifold of states. 
Due to the large bandwidth of the SNAIL mode, it is easy to create distinguishable photons unless great care is taken to match the two pump frequencies and phases to create identical excitation frequencies in the SNAIL. 
In this case, certain states of the qubit become ``dark'' to the parametric process. 
We have separately investigated this in a similar sample~\cite{ZHOU_Superconducting_2024}, where the drives with required phase stability are generated using high-speed RFSoC electronics and the QICK platform~\cite{STEFANAZZI_QICK_2022}.
In this work, however, we focus on distinguishable photons in our finite temperature bath, so that each ``heating''/``cooling'' event is uncorrelated with others.

\textit{Experiment results} - Our experiment scheme is shown in the lower panel of Fig.~\ref{fig:experimentdiagram}(a). 
The relevant circuit parameters are given in Appendix~\ref{app: circuit params and setup}.
A Traveling Wave Parametric Amplifier (TWPA)~\cite{MACKLIN_Quantum_2015} is used for single-shot readout and transmon state differentiation. 
We perform a three-part division on the readout data and label all states from $\ket{f}$ and beyond as $\ket{f^+}$.

\begin{figure}
\includegraphics{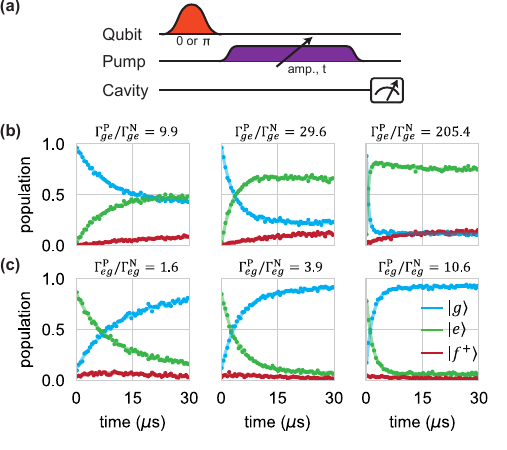}
\caption{\textbf{Result of heating and cooling processes.} 
(a) Pulse sequence of the experiment: The qubit is initialized in the ground/excited state for the ``heating''/``cooling'' experiment, respectively.
Next, we apply the corresponding pumps with varying amplitude and duration.
The qubit is measured at the end of the pump. 
(b) Qubit state populations (blue $\ket{g}$, green $\ket{e}$ and red $\ket{f^+}$) as a function of pump duration in the heating process with increasing pump strength $g_{\Sigma_{s, ge}}$, from left to right. 
(c) Qubit state populations as a function of pump duration in the cooling process with increasing pump strength $g_{\delta_{s, ge}}$. 
All experiment data are fitted to a semi-classical model (solid line, see Appendix~\ref{parametricbathengineering}) to extract the effective state transition rates.
}
\label{fig:separateprocesses}
\end{figure}

We begin with the implementation of the ``heating'' and ``cooling'' processes. 
In the ``heating'' experiment, the qubit is initialized in its $\ket{g}$ state. 
We apply the $\Sigma_{\text{s}, ge}$ drive while sweeping its strength and duration before the qubit measurement, as shown in Fig.~\ref{fig:separateprocesses}(a). 
The qubit state population as a function of the pump duration is shown in Fig.~\ref{fig:separateprocesses}.
The pump strength increases from left to right, resulting in a rise of the heating rate ($\Gamma^{\text{p}}_{ge}/\Gamma^{\text{N}}_{ge} = 9.9,~29.6,~205.4$ respectively; here p and N indicate the pumped and natural transition rates) as expected. 
The rate is extracted by fitting the data to the theory model (solid line, see Appendix~\ref{parametricbathengineering}).
We have observed a nontrivial $\ket{f^+}$ state population, attributed to the qubit's finite temperature, which excites it from $\ket{e}$ to $\ket{f}$. 
Additionally, a sufficiently strong pump may cause extra parametric heating due to its finite spectral weight on the transmon's higher transitions.
We emphasize that for strong heating drives, the qubit population inverts, rather than spreading to the higher states \textemdash~a scenario typically achieved by contacting it with a hot bath.

In the ``cooling'' experiment, the qubit is initialized in the $\ket{e}$ state. 
Then the $\delta_{s, ge}$ pump is applied to the system with various pump strengths and pump durations, as shown in Fig.~\ref{fig:separateprocesses}(a).
As shown in Fig.~\ref{fig:separateprocesses}(c), the transmon thermalizes to its equilibrium state faster as the pump strength increases ($\Gamma^{\text{p}}_{eg}/\Gamma^{\text{N}}_{eg} = 1.6,~3.9,~10.6$, respectively).
In contrast to~\cite{GEERLINGS_Demonstrating_2013} where both the qubit and the cavity are driven to force the qubit fully into the $\ket{g}$ state, our method allows us to determine the $\ket{g}$ probability and only requires one drive on the SNAIL. 
Unlike the heating experiment, there is no rise in the $\ket{f^+}$ state, because we are actively pushing the qubit to the ground state and there are no parametric processes that can bring the transmon to the $\ket{f^+}$ state at a frequency near the $\delta_{s, ge}$ pump. 

\begin{figure}
\includegraphics{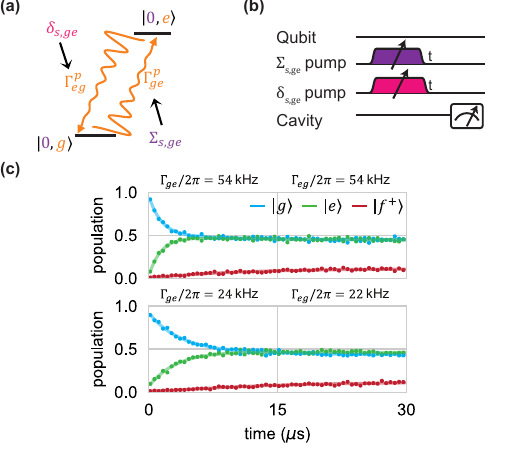}
\caption{\textbf{Heating and cooling balance.} 
(a) Schematic of the process. Both $\Sigma_{s, ge}$ and $\delta_{s, ge}$ pumps are applied to the system simultaneously to reach a $50\% \ket{g}$ - $50\% \ket{e}$ mixed state by balancing both state transition rates. 
(b) Pulse sequence of the experiment: The qubit is initialized in the ground state, followed by the $\Sigma_{s, ge}$ and $\delta_{s, ge}$ pumps with varying durations.
The qubit is measured at the end of the pumps.
(c) Qubit state populations as a function of time for two different pump settings.
Under different pump conditions, the qubit thermalizes to the same nearly-equal thermal mixture of $\ket{g}$ and $\ket{e}$ states with different rates. 
}
\label{fig:two-level both pump}
\end{figure}

The single pump experiments have demonstrated our control over each parametric process by establishing the equilibrium states with different effective temperatures.
We note that the difference in the final distribution for different single pump strengths does not conflict with Eq.~\ref{Eq: CP rho} for the case with $g_{\Sigma} = 0$ or $g_{\delta} = 0$, as the qubit also directly couples to the environment weakly in practice.

The full control of the bath is realized by applying both pumps simultaneously. 
As shown in Fig.~\ref{fig:two-level both pump}(a), when the $\Sigma_{s, ge}$ and $\delta_{s, ge}$ drives are both on, the two processes happen at the same time with different rates set by their respective pump strengths.
The system is initialized in the state $\ket{g, 0}$, and then both pumps are applied for a certain duration, followed by the qubit measurement (Fig.~\ref{fig:two-level both pump}(b)). 
In Fig.~\ref{fig:two-level both pump}(c), we show the qubit state population as a function of pump length for two sets of heating and cooling rates.
The pump conditions are chosen such that the ratio between the two rates is close to 1 ($\Gamma^{\text{p}}_{ge}/\Gamma^{\text{p}}_{eg} \sim 1$).
The qubit thermalizes to the same mixed $50\% \ket{g}$ - $50\% \ket{e}$ state (up to the residual population in the $\ket{f^+}$ state) with different rates. 
Additionally, by tuning the ratio between the two rates, we are able to set the transmon to a Gibbs state at an arbitrary temperature (see Fig.~\ref{fig:different_state_with_both_pumps}).

\begin{figure}[t]
\includegraphics{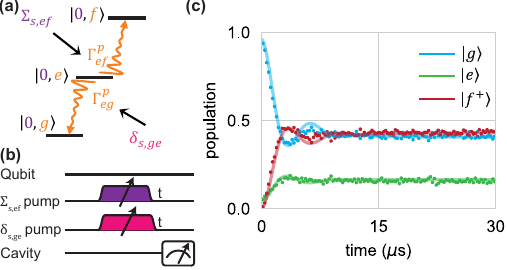}
\caption{\textbf{Bath engineering for a three-level qubit.} 
(a) Schematic of the process. Both $\delta_{s, ge}$ and $\Sigma_{s, ef}$ pumps are simultaneously applied with matched rates away from $\ket{e}$ to achieve a mixed state with $50\% \ket{g}$ - $50\% \ket{f^+}$. 
(b) Pulse sequence of the experiment: The qubit is initialized in the ground state, followed by the $\Sigma_{s, ef}$ and $\delta_{s, ge}$ pumps with varying durations.
The qubit is measured at the end of the pumps.
(c) Qubit state populations as a function of time under both pumps with the qubit initially prepared in $\ket{g}$. 
The $\ket{e}$ population is non-zero due to qubit decay from $\ket{f^+}$ and thermal excitations from $\ket{g}$.}
\label{fig:g_f_mix}
\end{figure}

Finally, we extend our protocol beyond the two-level manifold.
Similar to the two-level case, the Gibbs state of the three-level system with a tunable artificial temperature can be realized by properly driving the transition processes between different transmon energy levels (Appendix~\ref{general Gibbs state}). 
However, the flexibility of parametric control enables us to go beyond the Gibbs state. 
By applying external drives, we can engineer the diagonal elements of the steady-state density matrix, allowing the system to reach more exotic states with diverse population distributions.
Furthermore, if the pump tones are engineered to be mutually coherent, the off-diagonal elements can also be manipulated, enabling applications such as qubit longitudinal readout~\cite{ZHOU_Superconducting_2024} and synthetic squeezing~\cite{govia_stabilizing_2022}.

As a demonstration of the exotic steady state, we stabilize the transmon to a mixture of $50\% \ket{g}$ and $50\% \ket{f}$ (bottom right of Fig.~\ref{fig:experimentdiagram}(d)). 
This is achieved by simultaneously pumping the $\ket{e} \rightarrow \ket{g}$ cooling process at $\omega_{\delta_{s, ge}}$ and the $\ket{e} \rightarrow \ket{f}$ heating process at $\omega_{\Sigma_{s, ef}}$ with matched transition rates away from $\ket{e}$, as shown in Fig.~\ref{fig:g_f_mix}(a).
The transmon population as a function of pump duration is shown in Fig.~\ref{fig:g_f_mix}(c).
The residual $\sim 15\%$ population in the $\ket{e}$ state is due to the qubit decay from the $\ket{f}$ state and the thermal excitation from the $\ket{g}$ state.
Although the simple semiclassical model can predict the steady-state population distribution, it does not fully capture the dynamics of the measured data (Fig.~\ref{fig:3-level semi-classical}); we attribute this to the residual coherence between the two pump tones and the finite transmon thermalization rate elevated by the pumps. 
By including these factors, we are able to numerically simulate the system's behavior using QuTiP (Appendix~\ref{app:threelevelfitting}).

\textit{Discussion and Conclusion} - We present an engineered bath with a tunable temperature ranging from negative to positive values.
A key application of this bath is the Gibbs state preparation, as we demonstrate with the transmon.
Our approach can be extended to more general systems: given a bosonic system Hamiltonian $H$ and a set of its eigenoperators $A_i$, we show that the steady-state of the system will be the Gibbs state: $\rho = e^{-\beta H}/\text{Tr}(e^{-\beta H})$, if the system parametrically couples to lossy channels via its eigenoperators through the interactions similar to those in Eq.~\ref{Eq: tms eq} and~\ref{Eq: bs eq} (Appendix~\ref{general Gibbs state}). 

The ability to adjust bath temperature and the qubit's thermalization rate offers an opportunity for quantum simulation schemes across varied parameter regimes.
For example, when simulating energy transport in a spin chain~\cite{MEJIA-MONASTERIO_Heat_2007, YU_Theory_2011, ROY_Nonequilibrium_2024}, our system can be attached to different spin sites with a different effective temperature, providing tunable boundary conditions for the simulator. 
In particular, with the temperature tuning range from negative to positive, this bath can be viewed as a photon source/sink.
When combined with a large qubit array, it enables various simulations of quantum many-body systems, such as the investigation of Mott insulators~\cite{MA_Stabilizing_2019, MARK_Preparation_2012}, the study of entanglement and out-of-equilibrium steady state~\cite{DUTTA_Outequilibrium_2021, DU_Probing_2024}.
Moreover, access to the transmon's higher energy levels opens avenues for experimental studies of Gentile statistics~\cite{GENTILEJ._ItOsservazioni_1940, DAI_Gentile_2004, SRIVATSAN_Gentile_2006} and the simulation of condensed matter systems, such as magnons and spin waves~\cite{SHEN_Elementary_2022}. 

Going beyond the applications in quantum simulations, our system can be viewed as a single-cell open quantum battery from a quantum energy storage perspective~\cite{BARRA_Dissipative_2019}. 
The charging and discharging processes of the battery are enabled by engineered dissipation. 
Extending this setup to a multi-cell quantum battery by increasing the number of transmons is straightforward, which provides a platform to explore the charging power enhancement in collective charging schemes~\cite{FERRARO_HighPower_2018, CAMPAIOLI_Colloquium_2024}.
Additionally, the``heating" process forms a traditional lambda structure with a trapped excited state, offering a pathway to generate highly coherent microwave signals~\cite{LIU_Proposal_2021}.

\textit{Acknowledgment} - We acknowledge contributions to earlier versions of this experiment from Guanyu Zhu, Mohammad Hafezi, Mattias Fitzpatrick, Neereja Sundaresan, Basil Smithham, Christie Chou, and Andrew Houck. 
The TWPA used for amplified readout in this experiment was provided by MIT Lincoln Laboratory. 
This work was supported by the  Army Research Office under Grants No. W911NF15-1-0397 and M. Hatridge’s NSF CAREER grant (PHY-1847025). 
The views and conclusions contained in this document are those of the authors and should not be interpreted as representing official policies, either expressed or implied, of the Army Research Office or the U.S. Government. 
The US Government is authorized to reproduce and distribute reprints for government purposes not withstanding any copyright notation herein. 

\appendix

\section{Circuit parameters and experiment setup}
\label{app: circuit params and setup}
\begin{figure}[t]
\includegraphics{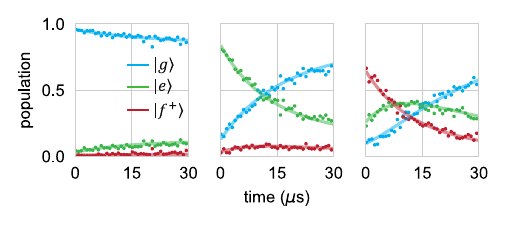}
\caption{\textbf{Transmon natural decay measurement.} 
The transmon is prepared in different initial states ($\ket{g}$, $\ket{e}$, and $\ket{f}$, shown from left to right) and allowed to thermalize to its natural steady state with no external pumps applied. 
The state populations are shown as a function of time, and the qubit's natural decay and heating rate can be obtained by fitting it to a semi-classical three-level model. 
}
\label{fig: natural decay} 
\end{figure}

The qubit frequency is $\omega_q/2\pi = 4.520~$GHz, anharmonicity $\alpha/2\pi = -197.39~$MHz, and its natural decay and heating rates are: $\Gamma^{\text{N}}_{ge}/2\pi = 1.22$~kHz, $\Gamma^{\text{N}}_{eg}/2\pi = 8.28$~kHz, $\Gamma^{\text{N}}_{ef}/2\pi = 2.20$~kHz, and $\Gamma^{\text{N}}_{fe}/2\pi = 12.27$~kHz. 
The rates are measured by initializing the qubit in different states (Fig.~\ref{fig: natural decay}). 
The SNAIL frequency (with no external flux applied) is $\omega_{s_{0}}/2\pi = 9.136~$GHz and its linewidth $\kappa_s/2\pi = 12.98~$MHz. 
To minimize the undesired higher-order effects of the SNAIL, we operate at its fourth-order Kerr cancellation point~\cite{SIVAK_KerrFree_2019}, at which the SNAIL frequency is $\omega_{s_{\text{op}}}/2\pi = 8.01~$GHz. 
The hybridization strength $\left(\frac{g}{\Delta}\right)_{sq} =  \num{1e-2}$, while $\left(\frac{g}{\Delta}\right)_{sc} =  \num{7e-4}$. 
The readout cavity has bandwidth $\kappa_c/2\pi = 0.761~$MHz and is followed by a Traveling Wave Parametric Amplifier (TWPA)~\cite{MACKLIN_Quantum_2015} provided by MIT Lincoln Laboratory through IARPA, which allows for single-shot readout and clear state differentiation.

\begin{figure*}
\includegraphics[scale=0.8]{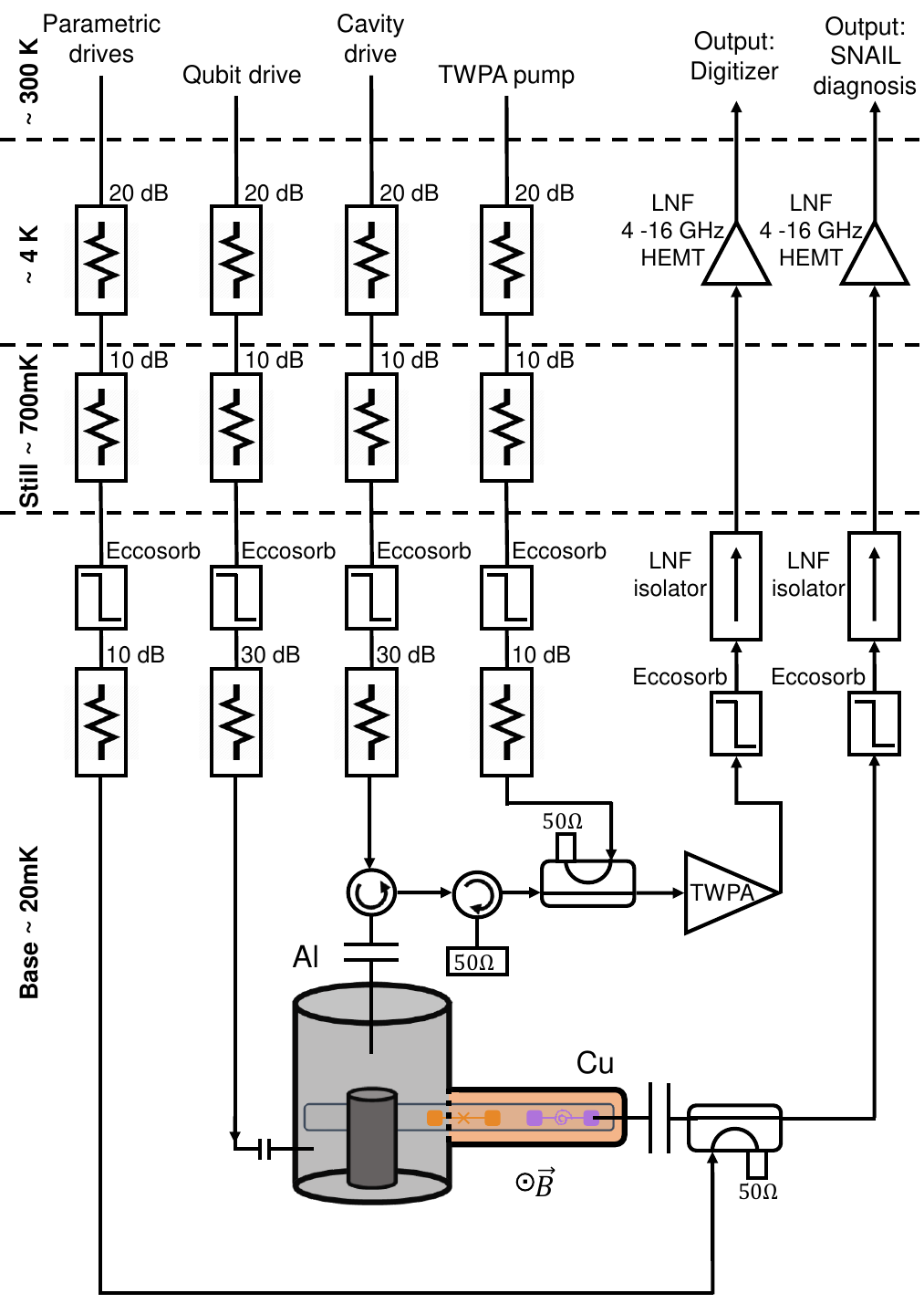}
\caption{\textbf{Cryogenic and wiring diagram}}
\label{fig:cryogenicdiagram} 
\end{figure*}

The entire experiment was cooled to approximately 20 mK on the base stage of a dilution refrigerator (Fig.~\ref{fig:cryogenicdiagram}). 
Each input drive line is attenuated and filtered with lossy filters made of Eccosorb CR-110 for high-frequency light. 
The external flux threaded through the SNAIL's superconducting Josephson junction loop is generated by DC signals of a current source (YOKOGAWA GS200) through superconducting wires.
Qubit control, parametric pump, and readout pulses are generated by mixing a local oscillator tone (generated by a SignalCore SC5511A rf signal generator) with pulses made by an arbitrary waveform generator (Tektronix AWG5014C) with a sampling rate of 1.0 GSa/s using an IQ-mixer (Marki IQ0618).

We measure the reflected signal phase shift from the readout cavity to dispersively infer the qubit's state ~\cite{BLAIS_Cavity_2004}. 
The reflected signal is amplified at the base stage by a TWPA, then moves through an Eccosorb filter to prevent high-frequency stray photons.
It next passes through microwave isolators on the output lines and is amplified using a Low Noise Factory HEMT amplifier at the 4K stage. 
Once outside of the fridge, the signal is mixed down to 50 MHz using an IR mixer (Marki IR4509) and a local oscillator signal from the SignalCore. The signal is then recorded using a fast ADC card (AlazarTech).

\begin{figure}
\includegraphics{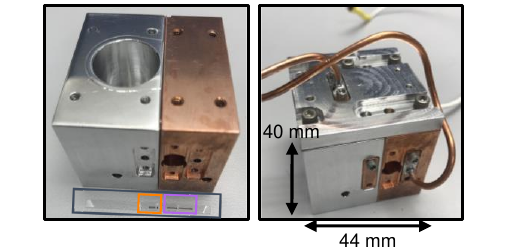}
\caption{\textbf{Photos of the device.} 
We use a two-part package in our experiment.
The aluminum piece is a High-Q coaxial microwave cavity used for measuring the qubit. 
The copper piece provides a tunnel-style chip holder that allows flux tuning of the SNAIL.
Both the transmon and SNAIL are fabricated on the same sapphire chip and is inserted into the aluminum cavity through the copper tunnel.}
\label{fig:devicephoto} 
\end{figure}

The sample housing, shown in Fig.~\ref{fig:devicephoto}, comprises two parts. 
The left part is a 6061 aluminum coaxial, $\lambda/4$ cavity~\cite{REAGOR_Reaching_2013} that acts as our readout resonator. 
An opening on the side of the coaxial cavity connects to an OFHC copper tube~\cite{AXLINE_Architecture_2016} that the sapphire spans to allow the flux biasing of the SNAIL.
The qubit and SNAIL are fabricated out of aluminum, with Al/AlOx/Al Josephson junctions, on the same chip of sapphire, via nanoscale lithography. 
The qubit is positioned between the aluminum and copper housing sections to optimize inter-element couplings.
Microwave coupling pins were inserted into the housing, one each per element. 

\section{SNAIL-qubit Hamiltonian}
\label{FullSystemHamiltonian}
The transmon-SNAIL Hamiltonian can be written as:
\begin{align}
    H_{0}\slash\hbar  &= \omega_\text{q} q^{\dagger}q + c^\text{q}_4 (q + q^{\dagger})^4 \nonumber \\
    &+ \omega_\text{s} s^{\dagger} s +  c^\text{s}_3(s + s^{\dagger})^3 \nonumber \\
    &+ g(s^{\dagger}q + s q^{\dagger}),
    \label{Eq:StaticCouplingHamiltonian}
\end{align}
where $\omega_\text{q}$ is the transmon frequency, $\omega_\text{s}$ is the SNAIL frequency, $g$ is the capacitive coupling strength between the transmon and SNAIL, $c^\text{q}_4$ is the coefficient of the transmon's fourth-order nonlinearity, and $c^\text{s}_3$ is the coefficient of the SNAIL's third-order nonlinearity. 
Note here we neglect the fourth-order term from the SNAIL mode, as it can be largely suppressed at its optimal working point. 
In addition, the Hamiltonian for the SNAIL pump is given by:
\begin{equation}
    H_{\text{drive}}  = \epsilon_\text{d} ( e^{i(\omega_\text{d} t + \phi_\text{d})}s^{\dagger} + e^ {-i(\omega_\text{d} t + \phi_\text{d})} s),
    \label{Eq:DrivingHamiltonian}
\end{equation}
where $\omega_\text{d}$, $\epsilon_\text{d}$, and $\phi_\text{d}$ are the driving frequency, amplitude, and phase respectively.

In the dispersive regime, i.e., $g \ll \Delta = \abs{\omega_\text{s} - \omega_\text{q}}$, the system can be re-diagonalized under the Bogoliubov transformation~\cite{BLAIS_Circuit_2021}.
Note that all operators and coefficients are transformed from their original basis, but marked as such for simplicity:
\begin{align}
    H_{\text{sys}}\slash\hbar  &=  \omega_\text{q} q^{\dagger}q + c^\text{q}_4 (q + \frac{g}{\Delta} s + \text{h.c.})^4 \nonumber \\
    &+ \omega_\text{s} s^{\dagger} s + c^\text{s}_3 (s + \frac{g}{\Delta} q + \text{h.c.} )^3 \nonumber \\
    &+ \epsilon_\text{d} ( e^{i(\omega_\text{d} t + \phi_\text{d})}s^{\dagger} + e^ {-i(\omega_\text{d} t + \phi_\text{d})} s)
    \label{Eq:DrivenHybridizedHamiltonian}
\end{align}
The drive terms can be removed by moving to a displacement frame with the transformation: $H' = O^{\dagger} H O - iO^{\dagger} \dot{O}$, where the transform operator is chosen to be $O_{\text{D}} = \text{exp}(\beta s^{\dagger} - \beta^*s)$, with $\beta = \epsilon_\text{d} e^{-i \phi_\text{d}} / (\omega_\text{d} - \omega_\text{s})$.
The system Hamiltonian is then given by:
\begin{align}
    H_{\text{sys}}\slash\hbar  &=  \omega_\text{q} q^{\dagger}q + c^\text{q}_4 (q + \frac{g}{\Delta} (s + \beta e^{i(\omega_\text{d} t + \phi_\text{d})}) + \text{h.c.})^4 \nonumber \\
    &+ \omega_\text{s} (s + \beta e^{i(\omega_\text{d} t + \phi_\text{d})})(s^{\dagger} + \beta^* e^{-i(\omega_\text{d} t + \phi_\text{d})}) \nonumber \\ 
    &+ c^\text{s}_3 (s + \beta e^{i(\omega_\text{d} t + \phi_\text{d})} + \frac{g}{\Delta} q + \text{h.c.})^3 
\end{align}
This can be further simplified by moving into a rotating frame in a similar fashion by choosing the transform operator to be $O_{\text{R}} = e^{i \omega_\text{s} s^{\dagger} s t + i \omega_\text{q} q^{\dagger} q t}$.
We get:
\begin{align}
H_{\text{sys}}\slash\hbar  &= 24 c^\text{q}_4 |\beta|^2 q^{\dagger} q - \alpha q^{\dagger}qq^{\dagger}q + c^3_\text{s}( s e^{-i \omega_\text{s} t} \nonumber \\
&+ \frac{g}{\Delta} q e^{-i \omega_\text{q} t} + \beta e^{-i \omega_\text{d} t} + \text{h.c.})^3.
    \label{Eq:DrivenRWAHamiltonian}
\end{align}
Here the first term is the qubit Stark shift due to the pump, and $\alpha = -6 c^\text{q}_4$ is the qubit anharmonicity.
The effects of the frequency-selected parametric pumps are shown in the third term.
Expanding this term in Eq.~\ref{Eq:DrivenRWAHamiltonian} produces all possible combinations of SNAIL and qubit operators; most of them can be ignored due to their fast rotations. 
The pump frequency provides a degree of freedom for selecting the desired processes by matching its frequency.
In particular, when the SNAIL is driven at $\omega_\text{d} = \omega_\text{s} + \omega_{q}$, the qubit-SNAIL Hamiltonian takes the form
\begin{align}
    H_{\Sigma}\slash\hbar &= 24 c^\text{q}_4 |\beta_{\Sigma}|^2 q^{\dagger} q - \alpha q^{\dagger}qq^{\dagger}q \nonumber \\
    &+ g_{\Sigma}(\beta_{\Sigma} s^{\dagger} q^{\dagger} + \beta_{\Sigma}^{*} sq),
\end{align}
where $g_{\Sigma} = 6 c^\text{s}_3 \frac{g}{\Delta}$. 
Likewise for the difference frequency pump $\omega_\text{d} = \omega_\text{s} - \omega_\text{q}$:
\begin{align}
    H_{\delta}\slash\hbar &= 24 c^\text{q}_4 |\beta_{\delta}|^2 q^{\dagger} q - \alpha q^{\dagger}qq^{\dagger}q \nonumber \\
    &+ g_{\delta}(\beta_{\delta} s^{\dagger} q + \beta_{\delta}^{*} sq^{\dagger}),
\end{align}
where $g_{\delta} = 6 c^\text{s}_3 \frac{g}{\Delta}$.
Both of these expressions feature pump-induced Stark shifting, the transmon's anharmonicity, and the respective effective two-body parametric process. 
Specifically for this application, we restrict the system to only neighboring transitions, allowing us to write Eqs.~\ref{Eq: sigma pump Hamiltonian} and~\ref{Eq: delta pump Hamiltonain}.

\section{Artificial temperature of a parametrically driven qubit}
\label{chempot_derivation}
\begin{figure}
\includegraphics{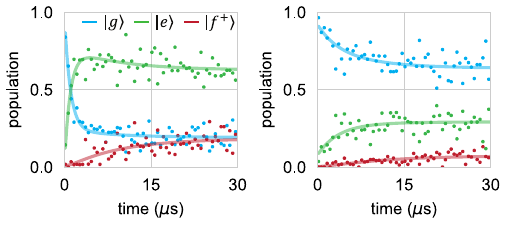}
\caption{\textbf{Qubit thermalization with different parametric drive amplitudes.}
The qubit state population as a function of time is shown for cases where $\Gamma_{ge} > \Gamma_{eg}$ (left) and $\Gamma_{ge} < \Gamma_{eg}$ (right).
}
\label{fig:different_state_with_both_pumps} 
\end{figure}

We next derive the expression for the artificial temperature of the qubit in the system shown in Fig.~\ref{fig:experimentdiagram}(a).
The system is modeled as follows: a two-level system (qubit) is parametrically coupled to a lossy linear mode (SNAIL), which is also in contact with a thermal bath at temperature $T_{\text{Bath}}$. 
As we discussed in the main text, when the two parametric processes are heralded by distinguishable photons due to the frequency detuning and/or phase instability between pump tones, no coherence will be established between them~\cite{ZHOU_Superconducting_2024}.
We therefore model them as two separate modes $s_1$ and $s_2$, similar to the treatment used in a non-degenerate amplifier~\cite{clerk_introduction_2010, frattini_optimizing_2018}. 
The Hamiltonian of the system is then given by:
\begin{align}
    H_{\text{sys}}/\hbar = &\frac{\omega_{q}}{2} \sigma^{z}_\text{q} + \omega_{s1} s_1^{\dagger} s_1 + \omega_{s2} s_2^{\dagger} s_2\nonumber \\ 
    &+ (g_\Sigma e^{i\omega_{\Sigma} t} s_1^{\dagger} \sigma^{+}_{\text{q}} + g_{\delta} e^{-i \omega_{\delta} t} s_2^{\dagger} \sigma^{-}_{\text{q}} + \text{h.c.}),
\end{align}
where $\omega_{s1}, \omega_{s2}$ are the two mode frequencies in proximity to the SNAIL's frequency: $|\omega_s - \omega_{s1, 2}| < \delta_p$, and $\delta_p$ is the frequency stability of the pump generator.
In the rotating frame, the Hamiltonian can be simplified as:
\begin{equation}
    H^{\text{R}}_{\text{sys}}/\hbar = g_\Sigma e^{i\omega_{\Sigma} t} s_1^{\dagger} \sigma^{+}_{\text{q}} + g_{\delta} e^{-i \omega_{\delta} t} s_2^{\dagger} \sigma^{-}_{\text{q}} + \text{h.c.}
\end{equation}
The master equation of this coupled system is given by:
\begin{align}
    \dot{\rho} = &\frac{i}{\hbar} [\rho, H^{\text{R}}_{\text{sys}}] \nonumber \\ 
    &+ \kappa_{\text{s}}(\Bar{N}(\omega_{\text{s}}) + 1) (D(s_1) \rho + D(s_2) \rho)  \nonumber \\
    &+ \kappa_{\text{s}} \Bar{N}(\omega_{\text{s}}) (D(s_1^{\dagger}) \rho + D(s_2^{\dagger}) \rho),
    \label{Eq: system master equation}
\end{align}
where $D(O)\rho = O \rho O^{\dagger} - \frac{1}{2}(\rho O^{\dagger} O + O^{\dagger} O \rho)$  is the Lindbladian superoperator.
Here we assume that only the SNAIL sees a thermal bath, and since $|\omega_s - \omega_{s1, 2}| \ll \omega_s$, the two modes are set to have the same thermal populations $\Bar{N}(\omega_{\text{s}}) = 1/(e^{\beta_{\text{B}} \hbar \omega_s} -1 )$, with $\beta_{\text{B}} = 1/k_B T_{\text{Bath}}$. 
Given that $\kappa_s \gg g_{\Sigma}, g_{\delta}$, the SNAIL mode can be adiabatically eliminated and we reach a master equation for the qubit mode:
\begin{align}
    \dot{\rho_q} = & \frac{4 g_{\Sigma}^2}{\kappa_s} [(\Bar{N}(\omega_{\text{s}}) + 1) D(\sigma_\text{q}^+) \rho + \Bar{N}(\omega_{\text{s}}) D(\sigma_\text{q}^-) \rho]   \nonumber \\
    &+ \frac{4 g_{\delta}^2}{\kappa_s} [(\Bar{N}(\omega_{\text{s}}) + 1) D(\sigma_\text{q}^-) \rho + \Bar{N}(\omega_{\text{s}}) D(\sigma_\text{q}^+) \rho].
    \label{Eq: rho dynamics}
\end{align}
The system density matrix at equilibrium is then given by solving the master equation with $\dot{\rho_q} = 0$:
\begin{equation}
    \rho_{\text{q}} (\infty) = \begin{pmatrix}
    \rho_{\text{q}}^{g} & 0 \\
    0 &\rho_{\text{q}}^{e} 
\end{pmatrix},
\end{equation}
with 
\begin{align}
    \rho_{\text{q}}^{g}  &= \frac{g^2_{\Sigma}\Bar{N}(\omega_{\text{s}}) +g^2_{\delta}(1 + \Bar{N}(\omega_{\text{s}})) }{(g^2_{\delta} + g^2_{\Sigma}) (1 + 2 \Bar{N}(\omega_{\text{s}})) }, \label{Eq: steady pg} \\
    \rho_{\text{q}}^{e} &= \frac{g^2_{\delta}\Bar{N}(\omega_{\text{s}}) +g^2_{\Sigma}(1 + \Bar{N}(\omega_{\text{s}}))}{(g^2_{\delta} + g^2_{\Sigma})(1 + 2 \Bar{N}(\omega_{\text{s}}))}. \label{Eq: steady pe}
\end{align}
Note that this is a thermal state $\rho_{\text{q}} = e^{-\beta_{\text{q}} \hbar \omega_{\text{q}} \sigma^z_{\text{q}}/2}/\text{Tr}(e^{-\beta_{\text{q}} \hbar \omega_{\text{q}} \sigma^z_{\text{q}}/2})$, with an inverse temperature:
\begin{equation}
    \beta_{\text{q}} = \frac{1}{\omega_{\text{q}}} \text{ln} (\frac{ g_{\Sigma}^2  + g_{\delta}^2 e^{\beta_{\text{B}} \omega_{\text{s}} } }{g_{\Sigma}^2 e^{\beta_{\text{B}} \omega_{\text{s}} } + g_{\delta}^2 })
    \label{Eq: inverse temperature two-level}
\end{equation}
We can thus tune the effective temperature of the qubit \textit{in situ}, by adjusting the ratio between the two processes' parametric rates, as shown in Fig.~\ref{fig:different_state_with_both_pumps}, as well as Fig.~\ref{fig:separateprocesses} in the main text.

\section{Dynamics of qubit thermalization}
\label{parametricbathengineering} 
\begin{figure}
\includegraphics{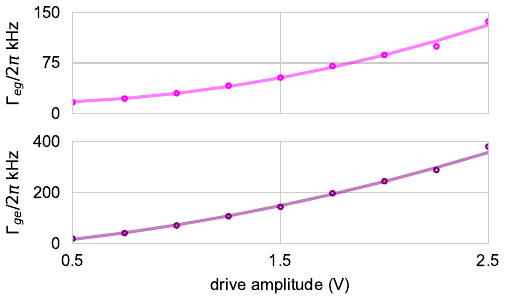}
\caption{\textbf{Effective ``heating'' and ``cooling'' rate.}
We show the effective rates of the parametric processes in the cooling (top) and heating (bottom) experiments as a function of room temperature drive voltage. 
The data fit nicely to a $g_{\Sigma,\delta}$ quadratic model as expected.} 
\label{fig:heatingcoolingstrengthcalibrateseparate} 
\end{figure}

To extract the effective ``heating'' and ``cooling'' rates, we need to understand the dynamics of the processes.
From Eq.~\ref{Eq: rho dynamics}, the state probability $P_g = \text{Tr}(\rho_{\text{q}} \ket{g}\bra{g})$ and $P_e = \text{Tr}(\rho_{\text{q}} \ket{e}\bra{e})$ are given by:
\begin{equation}
    \dot{P_g} = -\Gamma_{ge} P_g + \Gamma_{eg} P_e, 
\end{equation}
with the condition $P_{g} + P_{e} = 1$. 
The ``heating'' ($\Gamma_{ge}$) and ``cooling'' ($\Gamma_{eg}$) rates are given by:
\begin{align}
    \Gamma_{ge} &= \frac{4 g_{\Sigma}^2}{\kappa_s} (\Bar{N}(\omega_{\text{s}}) + 1) + \frac{4 g_{\delta}^2}{\kappa_s} \Bar{N}(\omega_{\text{s}})\\
    \Gamma_{eg} &= \frac{4 g_{\delta}^2}{\kappa_s} (\Bar{N}(\omega_{\text{s}}) + 1) + \frac{4 g_{\Sigma}^2}{\kappa_s} \Bar{N}(\omega_{\text{s}})
\end{align}
The solution is:
\begin{equation}
    P_g = c_0 e^{-(\Gamma_{ge} + \Gamma_{eg}) t} + \frac{\Gamma_{eg}}{\Gamma_{ge} + \Gamma_{eg}}, \label{Eq: semi-classical pg} 
\end{equation}
where $c_0$ is a constant to be determined by initial conditions. 
We note that the solution gives results that are consistent with Eqs.~\ref{Eq: steady pg} and~\ref{Eq: steady pe} at $t=\infty$.
To verify this result, we sweep the drive voltages in both the heating and cooling experiments and extract the corresponding rate. 
In Fig.~\ref{fig:heatingcoolingstrengthcalibrateseparate}, we show the heating and cooling rates as a function of the drive voltages. 
The data fit nicely to a quadratic function as predicted by the model.

For the cases where we also see a nontrivial $\ket{f^+}$ population, a similar but slightly more complicated model that involves the qubit's third level is used for obtaining the rate:
\begin{align}
    \dot{P_g} &= -\Gamma_{ge} P_g + \Gamma_{eg} P_e, \\
    \dot{P_e} &= \Gamma_{ge} P_g - (\Gamma_{ef} +  \Gamma_{eg}) P_e + \Gamma_{fe} P_f,
\end{align}
with $P_g + P_e + P_f = 1$.

\section{Numerical fitting of a three-level system}
\label{app:threelevelfitting} 

\begin{figure}
\includegraphics{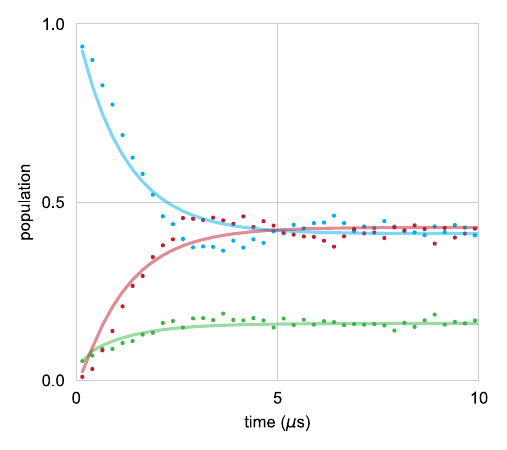}
\caption{\textbf{Three-level multi-pump data fit with semi-classical mode.}
The population of different states as a function of time. 
Here we zoom in on the first 10 $\mu$s of the data shown in Fig.~\ref{fig:g_f_mix}(b).
The data are fit to the semi-classical model introduced in Appendix~\ref{parametricbathengineering}. 
While the model can predict the steady-state population, it does not fully capture the dynamics of the entire process.
}
\label{fig:3-level semi-classical}
\end{figure}

Our model in the previous section agrees with the data when the pumps involve only the $\ket{g} \leftrightarrow \ket{e}$ manifold.
However, the dynamics of processes that involve pumps connected to all three levels cannot be accurately captured by it, as shown in Fig.~\ref{fig:3-level semi-classical}.
Here we go beyond the theoretical model and numerically study our system to understand its behavior in the three-level manifold. 
The effective Hamiltonian under the rotating wave approximation (in the frame rotating at $\omega_{ge}$) is:
\begin{align}
    H &= -\alpha q^{\dagger} q q^{\dagger} q +  (g_{\Sigma_{\text{s}, ef}} e^{i\alpha t} s^{\dagger} q^{\dagger} + \text{h.c.}) \nonumber \\
    &+ (g_{\delta_{\text{s}, ge}} s^{\dagger} q + \text{h.c.})
    \label{Eq: supplementFullHamiltonian}
\end{align}
We note that in the $\omega_{ge}$ frame, there will always be time-dependent terms that rotate at anharmonicity frequency $\alpha$ for the $\ket{e} \leftrightarrow \ket{f}$ manifold drives.
The full Lindblad master equation is given by:
\begin{equation}
    \dot{\rho} = -i[H, \rho] + \kappa_s D(s) \rho + \kappa^{\downarrow}_q D(q) \rho + \kappa^{\uparrow}_q D(q^{\dagger}) \rho.
\end{equation}
Here in the dissipator terms, we consider both the decay of the SNAIL and qubit (with collapse operator $s$ and $q$ at rate $\kappa_s$ and $\kappa^{\downarrow}_q$) and also the natural heating of the qubit due to the finite temperature (with collapse operator $q^{\dagger}$ at rate $\kappa^{\uparrow}_q$).
This equation is then simulated using QuTiP. 
We also take into account the up-pumped qubit thermalization during the measurement process.
This is modeled by turning off the pump tone in the Hamiltonian and letting the system evolve for the time of measurement.

The parameters used for the simulation we show in Fig.~\ref{fig:g_f_mix}(b) are as follows: $\alpha/2\pi = 197$ MHz, $\kappa_s/2\pi = 12.98$~MHz, $\kappa^{\downarrow}_q/2\pi = 0.029$~MHz, $\kappa^{\uparrow}_q/2\pi = 0.006$~MHz and measurement time $\tau_{msmt} = 1.2~\mu$s.
We note that the qubit decay rate used in the simulation is equivalent to a qubit with $T_1 \sim 4.5~\mu$s, which is faster than the transmon's natural decoherence rate.
We attribute this to the possible qubit coherence time reduction due to large pump photon numbers, which could invalidate the assumptions we used in the theoretical model.

\section{Gibbs state preparation via parametric bath engineering}
\label{general Gibbs state}
Our method of parametric bath engineering can be generalized to prepare Gibbs states for a general system. 
In this appendix, we show that, for a bosonic system Hamiltonian $H$ and a set of its eigenoperators $A_i$ that satisfy: $[H, A] = \lambda_i A_i$ and $\{A_i, A^{\dagger}_i, H \} = c I$, where $\{ \cdot \}$ denotes the commutant, $c$ is a constant and $I$ is the unit matrix.
We show that (as a sufficient condition) the steady state of the system will be the Gibbs state: $\rho_{\text{G}} = e^{-\beta H_{\text{sys}}}/\text{Tr}(e^{-\beta H_{\text{sys}}})$, if the system parametrically couples to lossy channels via its eigenoperators with the mode conversion and two-mode squeezing interactions. 
With the parametric interactions, the full system Hamiltonian is now:
\begin{equation}
    H = H_{\text{sys}} + \sum_i g_{\Sigma, i} A_i^{\dagger} B_{\Sigma, i}^{\dagger} + \sum_i g_{\delta, i} A_i^{\dagger} B_{\delta, i} + \text{h.c.},
\end{equation}
with the Lindblad master equation given by:
\begin{align}
    \dot{\rho} = &-i [H, \rho] \nonumber \\ 
    &+ \sum_i \kappa_{\Sigma, i} \left[ (N_{\Sigma, i} + 1) D(B_{\Sigma, i}) \rho + N_{\Sigma, i} D(B_{\Sigma, i}^{\dagger}) \rho \right] \nonumber \\
    &+ \sum_i \kappa_{\delta, i} \left[ (N_{\delta, i} + 1) D(B_{\delta, i}) \rho + N_{\delta, i} D(B_{\delta, i}^{\dagger}) \rho \right] , 
\end{align}
where the $B_{\Sigma, i}$ and $B_{\delta, i}$ are the lowering operators of the lossy modes that are associated with the corresponding parametric processes, $\kappa_{\Sigma, i}$ and $\kappa_{\delta, i}$ are their loss rates, and $N_{\Sigma, i}$ and $N_{\delta, i}$ are the thermal excitation occupation numbers of the corresponding baths. 
By adiabatically eliminating the lossy modes, we reach a master equation for the system of interest:
\begin{align}
    \dot{\rho}_{\text{sys}} = &-i [H_{\text{sys}}, \rho_{\text{sys}}] \nonumber \\  
    &+ \sum_i \frac{4 g_{\Sigma, i}^2 }{\kappa_{\Sigma, i}} \left[ (N_{\Sigma, i} + 1) D(A_i^{\dagger}) \rho_{\text{sys}} + N_{\Sigma, i} D(A_i) \rho_{\text{sys}} \right] \nonumber \\
    &+ \sum_i \frac{4 g_{\delta, i}^2 }{\kappa_{\delta, i}} \left[ (N_{\delta, i} + 1) D(A_i) \rho_{\text{sys}} + N_{\delta, i} D(A_i^{\dagger}) \rho_{\text{sys}} \right], 
\end{align}

To prove our statement, we first show that the Gibbs state is a steady state of the system, that is, $\rho_{\text{G}}$ is a solution to equation $\dot{\rho}_{\text{sys}} = 0$.
We then show the uniqueness of this solution to complete the proof. 

It is trivial to see that $[H_{\text{sys}}, \rho_{\text{G}}] = 0$, and since $A_i$ is the eigenoperator of the system Hamiltonian, we have $A_i \rho_{\text{G}} = e^{\beta \lambda_i} \rho_{\text{G}} A_i$.
This allows us to rewrite the second and third terms as:
\begin{widetext}
\begin{align}
    &\frac{4 g_{\Sigma, i}^2 }{\kappa_{\Sigma, i}} \left[ (N_{\Sigma, i} + 1) D(A_i^{\dagger}) \rho_{\text{G}} + N_{\Sigma, i} D(A_i) \rho_{\text{G}} \right] + \frac{4 g_{\delta, i}^2 }{\kappa_{\delta, i}} \left[ (N_{\delta, i} + 1) D(A_i) \rho_{\text{sys}} + N_{\delta, i} D(A_i^{\dagger}) \rho_{\text{G}} \right] \nonumber \\
    = & \left[\frac{4 g_{\Sigma, i}^2 }{\kappa_{\Sigma, i}} (N_{\Sigma, i} + 1) + \frac{4 g_{\delta, i}^2 }{\kappa_{\delta, i}} N_{\delta, i} \right] A_i^{\dagger} \rho_{\text{G}} A_i - \frac{1}{2} \left[\frac{4 g_{\Sigma, i}^2 }{\kappa_{\Sigma, i}} N_{\Sigma, i} + \frac{4 g_{\delta, i}^2 }{\kappa_{\delta, i}} (N_{\delta, i} + 1) \right] (A_i A_i^{\dagger} \rho_{\text{G}} + \rho_{\text{G}} A_i A_i^{\dagger}) \nonumber  \\
    &+ \left[\frac{4 g_{\delta, i}^2 }{\kappa_{\delta, i}} (N_{\delta, i} + 1) + \frac{4 g_{\Sigma, i}^2 }{\kappa_{\Sigma, i}} N_{\Sigma, i} \right] A_i \rho_{\text{G}} A_i^{\dagger} - \frac{1}{2} \left[\frac{4 g_{\delta, i}^2 }{\kappa_{\delta, i}} N_{\delta, i} + \frac{4 g_{\Sigma, i}^2 }{\kappa_{\Sigma, i}} (N_{\Sigma, i} + 1) \right] (A_i^{\dagger} A_i \rho_{\text{G}} + \rho_{\text{G}} A_i^{\dagger} A_i) \nonumber \\
    = & \left[\left(\frac{4 g_{\Sigma, i}^2 }{\kappa_{\Sigma, i}} (N_{\Sigma, i} + 1) + \frac{4 g_{\delta, i}^2 }{\kappa_{\delta, i}} N_{\delta, i} \right) - e^{\beta \lambda_i}\left(\frac{4 g_{\Sigma, i}^2 }{\kappa_{\Sigma, i}} N_{\Sigma, i} + \frac{4 g_{\delta, i}^2 }{\kappa_{\delta, i}} (N_{\delta, i} + 1) \right)  \right] A_i^{\dagger} \rho_{\text{G}} A_i \nonumber  \\
    &+ \left[\left(\frac{4 g_{\delta, i}^2 }{\kappa_{\delta, i}} (N_{\delta, i} + 1) + \frac{4 g_{\Sigma, i}^2 }{\kappa_{\Sigma, i}} N_{\Sigma, i} \right) - e^{-\beta \lambda_i}\left(\frac{4 g_{\delta, i}^2 }{\kappa_{\delta, i}} N_{\delta, i} + \frac{4 g_{\Sigma, i}^2 }{\kappa_{\Sigma, i}} (N_{\Sigma, i} + 1) \right)  \right] A_i \rho_{\text{G}} A_i^{\dagger}.
\end{align}
\end{widetext}
The $\rho_{\text{G}}$ will be the steady state under the following condition:
\begin{widetext}
\begin{equation}
    \left(\frac{4 g_{\Sigma, i}^2 }{\kappa_{\Sigma, i}} (N_{\Sigma, i} + 1) + \frac{4 g_{\delta, i}^2 }{\kappa_{\delta, i}} N_{\delta, i} \right) = e^{\beta \lambda_i}\left(\frac{4 g_{\Sigma, i}^2 }{\kappa_{\Sigma, i}} N_{\Sigma, i} + \frac{4 g_{\delta, i}^2 }{\kappa_{\delta, i}} (N_{\delta, i} + 1) \right), 
\end{equation}
\end{widetext}
and thus results in an artificial temperature of the system $\beta$:
\begin{equation}
    \beta = \frac{1}{\lambda_i} \text{ln} \left( \frac{\frac{4 g_{\Sigma, i}^2 }{\kappa_{\Sigma, i}} (N_{\Sigma, i} + 1) + \frac{4 g_{\delta, i}^2 }{\kappa_{\delta, i}} N_{\delta, i}}{\frac{4 g_{\Sigma, i}^2 }{\kappa_{\Sigma, i}} N_{\Sigma, i} + \frac{4 g_{\delta, i}^2 }{\kappa_{\delta, i}} (N_{\delta, i} + 1) }  \right).
    \label{Eq: general inverse temperature}
\end{equation}
When all the eigenoperators satisfy the condition: $\{A_i, A^{\dagger}_i, H \} = c I$, the uniqueness of the steady state is given by the Davies-Frigiero-Spohn criterion \cite{shabani_artificial_2016, FRIGERIO_Stationary_1978}. 
Note that Eq.~\ref{Eq: inverse temperature two-level} is a special case of Eq.~\ref{Eq: general inverse temperature} with $A_1 = \sigma^-_{\text{q}}$ and $H_{\text{sys}} = \frac{\omega_{\text{q}}}{2} \sigma^z_{\text{q}}$. 
And a similar relation for the three-level system can be obtained with $A_1 = \ket{e}\bra{g}$, $A_2 = \ket{f}\bra{e}$, and $H_{\text{sys}} = \sum_{i=g}^{f} \omega_{i} \ket{i}\bra{i}$. 

\bibliographystyle{apsrev4-1}
\bibliography{refs.bib}

\end{document}